\documentclass[12pt,preprint]{aastex}
\input epsf
\def\s2n{S^{\prime}/N}

\shorttitle{PRE--MAIN--SEQUENCE ACCRETION}
\shortauthors{PADOAN ET AL.}

\begin{document}
\title{A Solution to the Pre--Main--Sequence Accretion Problem}

\author{Paolo Padoan, Alexei Kritsuk and Michael L. Norman}
\affil{Department of Physics and Center for Astrophysics and Space Sciences, 
University of California, San Diego,\\
9500 Gilman Drive, La Jolla, CA 92093-0424; 
ppadoan@ucsd.edu, akritsuk@cosmos.ucsd.edu, mnorman@cosmos.ucsd.edu}
\author{\AA ke Nordlund}
\affil{Astronomical Observatory / NBIfAFG, Juliane Maries Vej 30, DK-2100, Copenhagen, Denmark; aake@astro.ku.dk}

\begin{abstract}

Accretion rates of order $10^{-8}$~M$_{\odot}$yr$^{-1}$ are observed in 
young pre--main--sequence (PMS) stars of approximately a solar mass with 
evidence of circumstellar disks. The accretion rate is significantly lower for
PMS stars of smaller mass, approximately proportional to the second 
power of the stellar mass, $\dot{M}_{\rm accr}\propto M^2$. 
The traditional view is that the observed accretion is the consequence 
of the angular momentum transport in isolated circumstellar disks, controlled 
by disk turbulence or self--gravity. However, these processes are not well understood 
and the observed accretion, a fundamental aspect of star formation, 
remains an unsolved problem. In this Letter we propose the stellar accretion 
rate is controlled by accretion from the large scale gas distribution in the 
parent cloud, not by the isolated disk evolution. Approximating this process 
as Bondi--Hoyle accretion onto the star--disk system, we obtain accretion 
rates comparable to the observed ones. We also reproduce the observed 
dependence of the accretion rate on the stellar mass. These results are 
based on realistic values of the ambient gas density and velocity, as 
inferred from numerical simulations of star formation in self--gravitating 
turbulent clouds.

\end{abstract}


\section{Introduction}

Accretion rates of order $10^{-8}$~M$_{\odot}$yr$^{-1}$ are observed in young 
pre--main--sequence (PMS) stars of mass, $M\sim 1$~M$_{\odot}$. Less massive
stars and brown dwarfs are also found to accrete, at a rate, 
$\dot{M}_{\rm accr}$, decreasing with the stellar mass, $M$, 
$\dot{M}_{\rm accr}\propto M^2$. A compilation of most accretion 
rates detected to date is shown in Figure~\ref{f1}. It includes 
PMS stars and brown dwarfs of $\approx 0.5$--$3$~Myr of age
\citep[][and references therein]{Natta+04,White+Hillenbrand04,Muzerolle+05}. 
The origin of such accretion rates and their dependence on the stellar 
mass is not understood. As accreting stars appear to be surrounded by 
circumstellar disks, it is usually suggested that the accretion is due to the 
viscous evolution of the disks, controlled by some form of ``turbulent viscosity''. 
The angular momentum transport necessary to match the observed accretion rate 
may result from self--gravity \citep[e.g.][]{Larson89,Laughlin+Bodenheimer94}, 
magneto--hydrodynamic turbulence \citep[e.g.][]{Gammie96,Stone+00}, or hydrodynamic
turbulence \citep[e.g.][]{Klahr+Bodenheimer03}. However, circumstellar disks
around PMS stars of few Myr of age contain only a few percent 
of the stellar mass and have a very low fractional ionization. 
As a consequence, both self--gravity and magnetic coupling may be 
ineffective, leaving perhaps only baroclinic instability as a viable 
source of turbulence \citep[see][and references therein]{Klahr+Bodenheimer03}.

Given the uncertainty in the source of disk turbulence and our limited understanding
of turbulent transport mechanisms, it is possible to compare the predicted
``viscous'' disk evolution with the observations only assuming ad hoc values
of the viscosity parameter \citep{Hartmann+98}. Furthermore, PMS stars 
are always found in association with their parent clouds. For 
example, most PMS stars in Taurus are confined within filaments of dense 
gas \citep{Hartmann02}. Therefore, we should expect an interaction of PMS stars 
and their circumstellar disks with the surrounding gas, and the assumption of an 
isolated disk evolution may be incorrect.

In this work we find that accretion from the large scale may be important, 
and may even account for most of the observed accretion rate. For the purpose
of discussion, we propose to consider two main phases of PMS evolution:
i) The formation phase and ii) the large scale post--formation accretion phase. The formation 
phase is roughly a spherical accretion. It ends because the initial mass reservoir 
is limited by the initial conditions (a more or less distinct turbulent fragment). 
It results in a contracting central object, the PMS star, and a circumstellar disk.
As the disk is initially massive, it is very unstable and rapidly accretes most of its 
mass onto the central object. After less than 1~Myr the circumstellar disk contains
only a few percent of the mass of the system, and the PMS star enters the second phase. 
The second phase is characterized by mass accretion from the large scale gas 
distribution in the parent cloud. This large scale accretion may account for 
most of the estimated protostellar accretion rates. It interacts with the 
circumstellar disk and may be an important source of disk turbulence, enhancing 
or even controlling the disk accretion. 

In this Letter we show that the large scale accretion, approximated as 
Bondi--Hoyle accretion mediated by the circumstellar disk, reproduces the 
observed accretion rate and its dependence on the stellar mass.

\section{Bondi--Hoyle Approximation for Large Scale Accretion}

We propose a solution to the PMS accretion problem that avoids the necessity
of an effective angular momentum transport mechanism in isolated disks. 
The basic idea is that young accreting PMS stars are not 
isolated objects. Even if they do not appear to be heavily embedded in 
protostellar cores, they are still associated with their parent molecular 
cloud and the accretion from the large scale gas distribution in the cloud
cannot be neglected.

The initial formation phase of a star is essentially a spherical accretion.
In this phase we assume that self--gravity, shocks, turbulence and magnetic fields 
in combination are able to effectively transfer angular momentum outwards. Due to this 
effective angular momentum transport, a central object containing most of the mass is 
formed, surrounded by a rotationally supported disk of smaller mass.
The angular momentum problem may be more serious when the disk contains only 
a few percent of the protostellar mass and if the ionization fraction is very low.
In this case, self--gravity and magnetic fields may not be responsible for the 
angular momentum transport, and some alternative source of turbulence is required 
\citep[see][and references therein]{Klahr+Bodenheimer03}. However, even if 
disk accretion were almost interrupted due to the lack of angular momentum 
transport in an isolated disk, accretion from the large scale flow of the parent 
cloud is unavoidable. This large scale accretion may represent the
main mass reservoir for the observed PMS accretion. The accreting
material from the large scale is likely to be reprocessed through the disk,
and may also be responsible for ``activating'' the disk accretion it fuels, by
destabilizing the disk (e.g. generating disk turbulence).

The accretion from the large scale lacks the spherical symmetry of the
initial formation phase, because it occurs far from hydrostatic equilibrium.
The gas flow is initially along the dense filaments where the stars are born, 
and later originates from the general turbulent flow.
In order to estimate the accretion rate, we assume this process can be 
approximated as Bondi--Hoyle accretion. In the idealized Bondi--Hoyle 
accretion, the star is in supersonic motion with respect to the gas. 
The stellar gravitation focuses the stream of gas and generates a shock 
at the back side of the star. In this shock, excess angular momentum with 
respect to the star cancels, and the gas can fall onto the star from the back 
side. The Bondi--Hoyle radius of this accretion flow that eventually falls onto 
the star spans the range 10--1000~AU for a star--to--gas velocity of
1~km/s and stellar masses in the range 0.01--1~M$_{\odot}$ respectively. 
This idealized Bondi--Hoyle flow may thus interact with the circumstellar disk 
(radius of order 100~AU), particularly for small stellar masses. 

In a more realistic description of a star--forming cloud, the accreting material
is not uniform and carries a finite amount of angular momentum, due to the 
turbulent nature of the ambient medium. \cite{Krumholz+05} have found that 
angular momentum may reduce the Bondi--Hoyle accretion rate onto the star.
However, most material should accrete onto the star--disk system. 
Numerical simulations have shown that, even in the presence of angular momentum, 
the accretion rate into a central region of size as small as one percent of the 
Bondi--Hoyle radius (hence smaller than the circumstellar disk radius) is 
generally comparable to the Bondi--Hoyle estimate \citep{Ruffert+Anzer95,Ruffert97,
Ruffert99}. Excess angular momentum with respect to the star may cancel in shocks 
with the disk. Details depend on the orientation of the accretion flow relative 
to the disk and on the size of the Bondi--Hoyle radius relative to the disk radius. 
We use the Bondi--Hoyle accretion rate, $\dot{M}_{\rm accr}\approx \dot{M}_{\rm BH}$,
as an order--of--magnitude estimate of the accretion from the large scale onto the 
star--disk system, not directly onto the star. The rate is given by
\begin{equation}
\dot{M}_{\rm BH}=\frac{4 \pi G^2 \rho_{\infty}}
                 {(c_{\infty}^2+v_{\infty}^2)^{3/2}} M^2
\label{eq1}
\end{equation} \\
\citep[][and references therein]{Bondi+Hoyle44,Bondi52,Edgar04}.
Here, $\rho_{\infty}$, $c_{\infty}$ and $v_{\infty}$ are the gas density,
sound speed and gas velocity relative to the star, at a large distance.

Assuming an average gas density $\rho_{\infty}=2.46m_{\rm H}10^3$~cm$^{-3}$
($m_{\rm H}$ is the mass of the proton), 
a velocity of the gas relative to the star $v_{\infty}=1$~km/s, 
a sound speed $c_{\infty}=0.2$~km/s and a stellar mass $M= 1$~M$_{\odot}$, 
equation~(\ref{eq1}) yields $\dot{M}_{\rm accr}\approx 10^{-8}$~M$_{\odot}$yr$^{-1}$, 
as observed. Furthermore, as $\rho_{\infty}$, $c_{\infty}$ and $v_{\infty}$
do not depend on the stellar mass, the accretion rate is proportional 
to the second power of the stellar mass, $\dot{M}_{\rm accr}\propto M^2$, 
also consistent with the observations. For a brown dwarf
of 0.03~M$_{\odot}$, for example, the above physical parameters give
$\dot{M}_{\rm accr}\approx 10^{-11}$~M$_{\odot}$yr$^{-1}$, consistent with
inferred accretion rates for young brown dwarfs. 

The values adopted above for $\rho_{\infty}$, $c_{\infty}$ and $v_{\infty}$
characterize molecular cloud filaments and vary independently of the stellar mass.
Their variation is thus a natural cause for the scatter in the observed accretion 
rates visible in Figure~\ref{f1}, on top of the general trend. 

On the scale of a few parsecs, jets and outflows from PMS stars are 
not the primary source of the turbulence that controls the large scale 
accretion, as indicated by statistical studies of turbulence in 
star--forming clouds \citep{Ossenkopf+MacLow02,Heyer+Brunt04}.
However, on smaller scale, jets and outflows may directly interact 
with the accretion flow \citep[eg][]{Arce+Sargent04} and may be 
another source of the scatter in the observed accretion rates. 
This is an important topic that should be addressed in future work.

\section{Large Scale Accretion in Turbulent Clouds}

Because stars are formed in dense filaments, and inherit their initial 
velocity from the gas in those filaments, the initial large scale accretion 
rate is always high in young PMS stars ($\lesssim 1$~Myr), and can easily 
account for the upper envelope of the observational plot in Figure~\ref{f1}. 
However, at an age of a few Myr, stellar positions and velocities 
are not expected to be well correlated with the original filaments, 
and the large scale accretion rate is expected to decrease with time on the 
average. The vertical scatter seen in Figure~\ref{f1} is indeed attributed 
partly to an age dependence of the accretion rate \citep[e.g.][]{Natta+04}. 

To estimate the large scale accretion rate, we use a numerical simulation of 
star formation in self--gravitating turbulent clouds. We employ a 3--D parallel 
structured adaptive mesh refinement (AMR) code, {\em Enzo}
\footnote{See {\tt http://cosmos.ucsd.edu/enzo/}},
developed at the Laboratory for Computational Astrophysics by Bryan,
Norman and collaborators \citep{bryan.99,oshea......04}.
In this simulation, mesh refinement is automatically carried out in collapsing 
regions in order to always resolve the Jeans length \citep{truelove.....97}. 
We solve the hydrodynamic equations (the case of magnetized clouds will be 
presented in a separate work), including self--gravity and a large scale 
random force to drive the turbulence. We also adopt an isothermal equation of 
state and periodic boundary conditions. The AMR allows us to achieve a very 
large dynamical range. The computational box has a size of 5~pc and the 
gravitational collapse is resolved down to the scale of 2~AU. The experiment 
and the code will be described in detail elsewhere.

\begin{figure*}
\epsscale{1.} 
\centerline{
\plotone{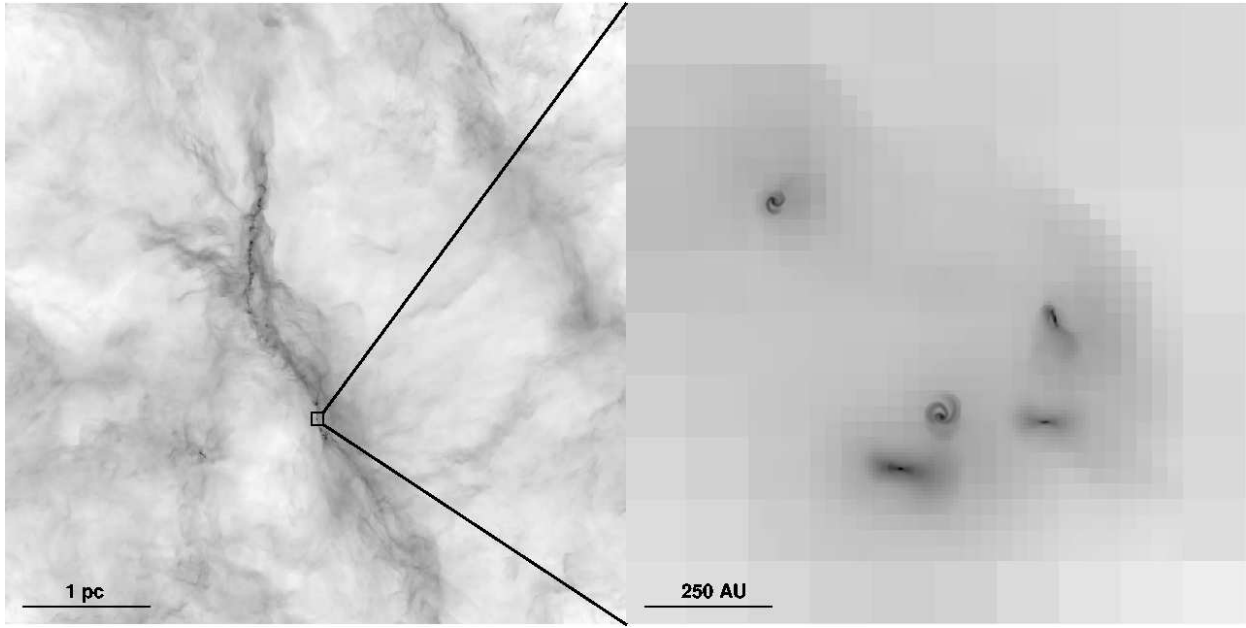}
}
\caption{Left panel: Logarithm of the projected gas density in our AMR simulation
of star formation in a turbulent cloud (see text). Right panel: Higher resolution
view of a small region inside the dense filament marked by the small square in the 
left panel. The small square is drawn larger than its actual size, as the 
magnification factor between the right and the left panels is 840.}
\label{f2}
\end{figure*}

In this simulation, a very complex density field is generated by the supersonic 
turbulence. Highly fragmented dense filaments are the natural result of the turbulent 
fragmentation process. The densest fragments in the filaments are 
gravitationally unstable and collapse into protostars. The left panel of 
Figure~\ref{f2} shows a typical filament found in the simulation, where several 
protostars are formed. The protostars appear in the filament as very dense cores, 
because of the limited dynamical range of the image, but are actually resolved 
as central objects surrounded by protostellar disks at higher refinement levels, 
as shown in the right panel of Figure~\ref{f2}.

We can use our experiment to derive the gas density and velocity distributions in 
the turbulent medium and infer the average and standard deviation 
of the large scale accretion rate based on the Bondi--Hoyle formula.  
We interpret this as an estimate of the accretion rate for PMS stars 
of a few Myr of age, assuming i) stellar velocities and positions are 
not correlated with the gas after a few Myr from their formation, 
and ii) the parent cloud has not been (completely) dispersed yet. 

We compute the value of the coefficient of equation (\ref{eq1}), 
$4 \pi G^2 \rho_{\infty}/(c_{\infty}^2+v_{\infty}^2)^{3/2}$,
at each mesh position, adopting the local values of gas density,
sound speed and flow velocity for $\rho_{\infty}$, $c_{\infty}$ 
and $v_{\infty}$ respectively. The temperature is uniform, $T=10$~K,
and the sound speed is constant. The mean density is 500~cm$^{-3}$
and the rms flow velocity is approximately 1.1~km/s, typical of molecular 
clouds on the scale of 5~pc. It is important to compute directly 
the coefficient of equation (\ref{eq1}), and not the density and velocity
separately, because the density and flow velocity in supersonic turbulence
are known to be correlated \citep{Padoan+01cores}. 
Averaging over the whole computational volume, we find 
$\dot{M}_{\rm accr}=10^{-8.6\pm 0.8}(M/{1~\rm M_{\odot}})^2~{\rm M_{\odot}yr^{-1}}$. 
As explained above, this should provide an order--of--magnitude estimate 
for the accretion rate of PMS stars of a few Myr of age,
whose velocities and positions are no longer correlated with the gas in dense
filaments. Indeed, the $\pm$1--$\sigma$ values of this accretion rate covers
approximately the lower half of the plot in Figure~\ref{f1} (dashed lines).  

As an illustration of the large scale accretion rate for younger PMS stars
($\lesssim 1$~Myr) formed in a turbulent cloud, we can compute the same 
coefficient from the simulation, but averaging only in regions with 
density above a certain value. This would mimic the fact that stars are 
initially born inside dense filaments. In principle, increasing values 
of the minimum density should correspond to lower ages. As an example, 
the accretion rate is 
$\dot{M}_{\rm accr}=10^{-7.6\pm 0.7}(M/{1~\rm M_{\odot}})^2~{\rm M_{\odot}yr^{-1}}$,
limited to densities above $10^3$~cm$^{-3}$, and
$\dot{M}_{\rm accr}=10^{-6.6\pm 0.6}(M/{1~\rm M_{\odot}})^2~{\rm M_{\odot}yr^{-1}}$,
for densities above $10^4$~cm$^{-3}$.
The solid lines in Figure~\ref{f1} show the $\pm$1--$\sigma$ values for
densities above $10^4$~cm$^{-3}$. These values can explain the 
upper envelope of the observational plot as due to large scale accretion 
on very young PMS stars. A self--consistent calculation of the large scale 
accretion rate of PMS stars formed in turbulent clouds will be the subject
of future work.

\section{Conclusions}

We have shown that accretion from the large scale gas distribution in the
parent cloud can explain the observed PMS accretion rates and their
dependence on the stellar mass. This process is approximated as Bondi--Hoyle
accretion onto the star--disk system in a turbulent molecular cloud and is 
distinct from the usual disk accretion, which assumes an isolated disk 
evolution. The large scale accretion provides both the mass reservoir for
the observed accretion and the possibility of destabilizing the disk
and ``activating'' its turbulent transport.
Regardless of how accretion is mediated on the smallest scales,
it may be controlled by the large scale accretion from the ambient medium. 

An important consequence of this new picture is that the disk lifetime could 
be longer than the ratio of the disk mass and the accretion rate, 
because the disk is not the only mass reservoir for the accretion;
rather it is a sort of buffer between the large scale flow and the final
accretion onto the star. PMS accretion continues until the large scale accretion
is significantly diminished by an increased stellar velocity relative to 
the gas or is interrupted by the cloud dispersion. It may also be 
interrupted if the star ends up in a region of very low gas density.  

This large scale accretion may play an important role in the process of 
planet formation, because of the induced turbulence in disks that may
otherwise be stable. Turbulence may be important because it could speed
up the growth of grains, which is usually considered essential for the
origin of rocky planets, or because it may trigger or modify direct
gravitational collapse.

\begin{figure*}
\centerline{
\includegraphics[width=12cm]{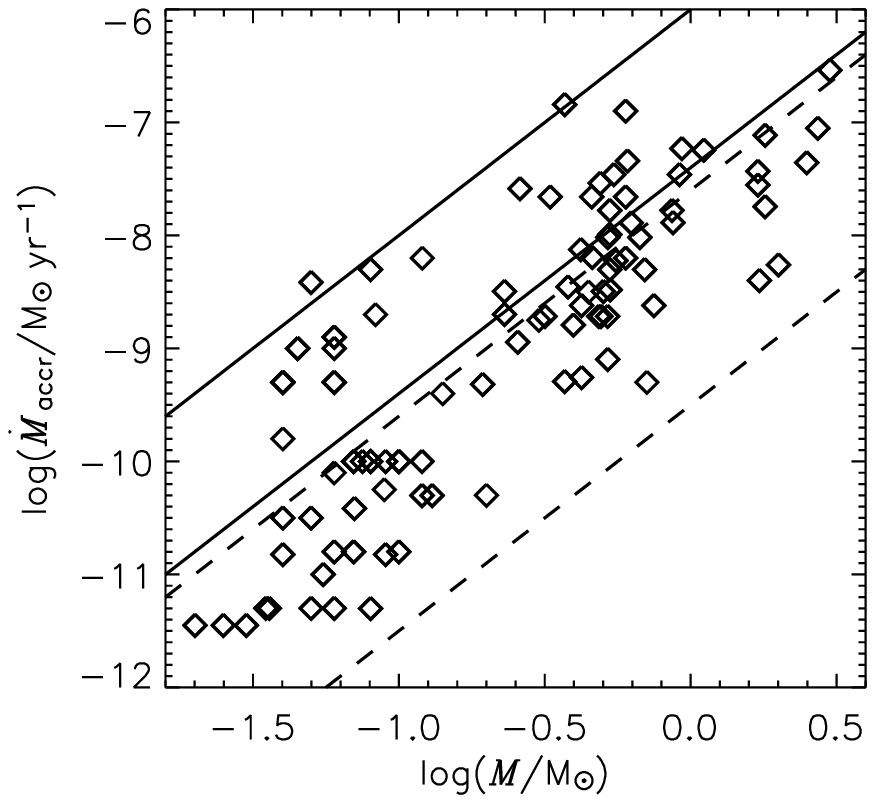}
}
\caption{Detected accretion rate versus stellar mass including 
all detections in \cite{Natta+04}, \cite{White+Hillenbrand04}, and 
\cite{Muzerolle+05} for pre--main--sequence stars and brown dwarfs
of 0.5--3~Myr of age. No upper limit has been included in this plot. The 
two solid lines show the 1--$\sigma$ scatter around the $M^2$ dependence 
predicted in \S~3 for young PMS stars ($\lesssim 1$~Myr), assuming a minimum 
density of $10^4$~cm$^{-3}$ for the accreting gas. The two dashed lines show 
the 1--$\sigma$ scatter around the $ M^2$ dependence predicted in \S~3 for older 
PMS stars (few Myr). The lower dashed line roughly corresponds also to the detection 
limit.}
\label{f1}
\end{figure*}

\acknowledgements
During the October 2004 brown dwarf conference in Volterra, PP enjoyed 
stimulating discussions on PMS accretion with Francesco Palla, 
Antonella Natta and Leonardo Testi, which inspired this work. James Muzerolle
and Russel White provided the accretion--rate data plotted in Figure~\ref{f1}.
Comments from the referee and from Lee Hartmann, Mark Krumholz, Russel White, Lynne 
Hillenbrand, James Muzerolle, Alyssa Goodman and John Scalo have helped
improve this Letter. The numerical simulation was done using the IBM 
Data Star system at the San Diego Supercomputer Center with support 
from NRAC award MCA098020S.


\end{document}